# The Nature of the Chemical Process. 1. Symmetry Evolution – Revised Information Theory, Similarity Principle and Ugly Symmetry

**Shu-Kun Lin**

Molecular Diversity Preservation International (MDPI), Sangergasse 25, Basel CH-4054 Switzerland. Tel. ++41 79 322 3379, Fax: ++41 61 302 8918, E-mail: lin@mdpi.org, http://www.mdpi.org/lin

---

**Abstract:** Symmetry is a measure of indistinguishability. Similarity is a continuous measure of imperfect symmetry. Lewis' remark that “gain of entropy means loss of information” defines the relationship of entropy and information. Three laws of information theory have been proposed. Labeling by introducing nonsymmetry and formatting by introducing symmetry are defined. The function $L$ ( $L=\ln w$, $w$ is the number of microstates, or the sum of entropy and information, $L=S+I$) of the universe is a constant (the first law of information theory). The entropy $S$ of the universe tends toward a maximum (the second law law of information theory). For a perfect symmetric static structure, the information is zero and the static entropy is the maximum (the third law law of information theory). Based on the Gibbs inequality and the second law of the revised information theory we have proved the similarity principle (a continuous higher similarity–higher entropy relation after the rejection of the Gibbs paradox) and proved the Curie-Rosen symmetry principle (a higher symmetry–higher stability relation) as a special case of the similarity principle. The principles of information minimization and potential energy minimization are compared. Entropy is the degree of symmetry and information is the degree of nonsymmetry. There are two kinds of symmetries: dynamic and static symmetries. Any kind of symmetry will define an entropy and, corresponding to the dynamic and static symmetries, there are static entropy and dynamic entropy. Entropy in thermodynamics is a special kind of dynamic entropy. Any spontaneous process will evolve towards the highest possible symmetry, either dynamic or static or both. Therefore the revised information theory can be applied to characterizing all kinds of structural stability and process spontaneity. Some examples in chemical physics have been given. Spontaneous processes of all kinds of molecular interaction, phase separation and phase transition, including symmetry breaking and the densest molecular packing and crystallization, are all driven by information minimization or symmetry maximization. The evolution of the universe in general and evolution of life in particular can be quantitatively considered as a series of symmetry breaking processes. The two empirical rules – similarity rule and complementarity rule – have been given a theoretical foundation. All kinds of periodicity in space and time are symmetries and contribute to the stability. Symmetry is beautiful because it renders stability. However, symmetry is in principle ugly because it is associated with information loss.

**Key words**: continuous symmetry, static entropy, similarity principle, symmetry principle, the second law of information theory, structural stability, complementarity, order, disorder

---

## 1. Introduction

Symmetry has been mainly regarded as a mathematical attribute [1-3]. The Curie-Rosen symmetry principle [2] ] is a higher symmetry–higher stability relation that has been seldom, if ever, accepted for consideration of structural stability and process spontaneity (or process irreversibility). Most people accept the higher symmetry–lower entropy relation because entropy is a degree of disorder and symmetry has been erroneously regarded as order [4]. To prove the symmetry principle, it is necessary to revise information theory where the second law of thermodynamics is a special case of the second law of information theory.

Many authors realized that, to investigate the processes involving molecular self-organization and molecular recognition in chemistry and molecular biology and to make a breakthrough in solving the outstanding problems in physics involving critical phenomena and spontaneity of symmetry breaking process [5], it is necessary to consider information and its conversion, in addition to material, energy and their conversions [6]. It is also of direct significance to substantially modify information theory, where three laws of information theory will be given and the similarity principle (entropy increases monotonically with the similarity of the concerned property among the components (Figure 1) [7]) will be proved.

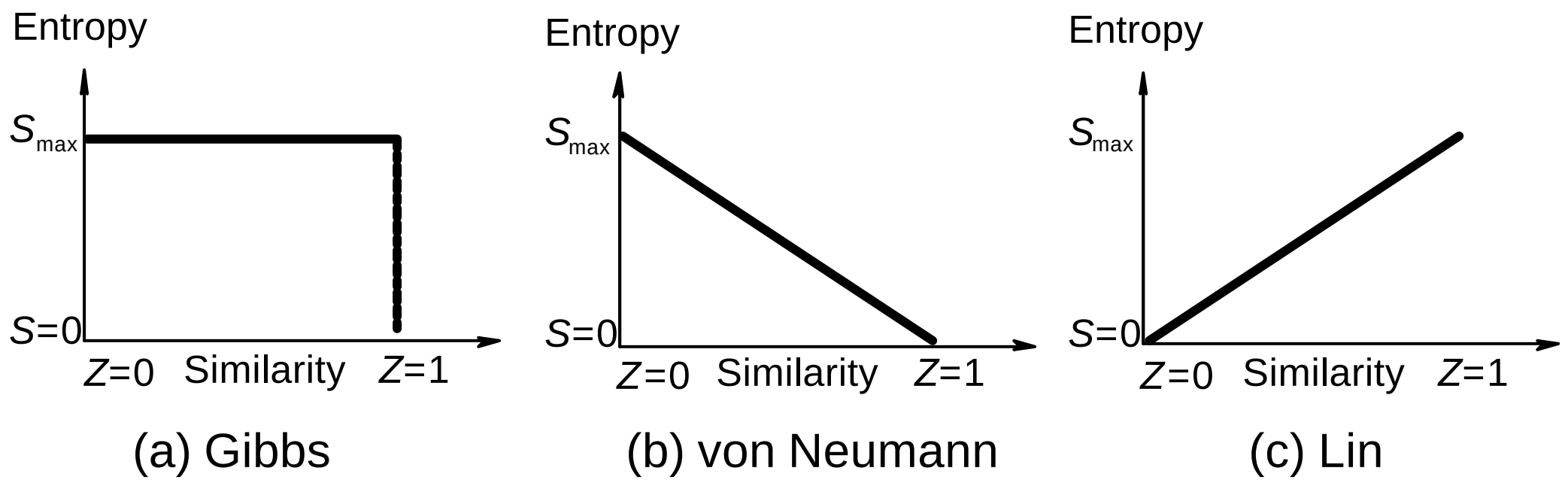

**Figure 1.** (a) Correlation of entropy (ordinate) of mixing with similarity (abscissa) according to conventional statistical physics, where entropy of mixing suddenly becomes zero if the components are indistinguishable according to the Gibbs paradox [21]. Entropy *decreases discontinuously*. Figure 1a expresses Gibbs paradox statement of "same or not the same" relation. (b) von Neumann revised the Gibbs paradox statement and argued that the entropy of mixing *decreases continuously* with the increase in the property similarity of the individual components [21a,21b,21d,21j]. (c) Entropy *increases continuously* according to the present author [7] (not necessarily a straight line because similarity can be defined in different ways).

Thus, several concepts and their quantitative relation are set up: higher symmetry implies higher similarity, higher entropy, less information and less diversity, while they are all related to higher stability. Finally, we conclude that the definition of symmetry as order [4] or as "beauty" (see: p1 of ref. 8, also ref. 3) is misleading in science. Symmetry is in principle ugly. It may be related to the perception of beauty only because it contributes to stability.

## 2. Definitions

### *2.1. Symmetry and Nonsymmetry*

Symmetry as a Greek word means *same measure* [1]. In other words, it is a measure of indistinguishability. A number $w_S$ can be used to denote the measure of the indistinguishability and can be called the symmetry number [7d]. In some cases it is the number of invariant transformations. Clausius proposed to name the quantity $S$ the entropy of the body, from the Greek word η τροπη, a *transformation* [9]. This may suggest that symmetry and entropy are closely related to each other [7d].

Only perfect symmetry can be described mathematically by group theory [2]. It is a special case of imperfect symmetry or continuous symmetry, which can be measured by similarity, instead of indistinguishability [7d].

*Example 1.* The bilateral symmetry of the human body is imperfect. We have our heart on the left side. Detailed features are only similar on the two sides. The two breasts on the left and the right side of women are normally somewhat different in size and shape. The beautiful woman Cindy (www.cindy.com) has a black birthmark on the left side of her otherwise very symmetric face.

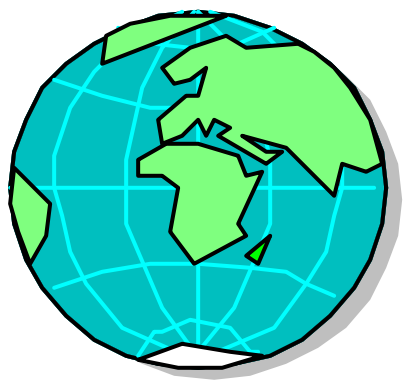

**Figure 2.** Our earth is not of perfect symmetry.

On the contrary, we define nonsymmetry as a measure of difference or distinguishability. Similarity can be defined as a continuous index of imperfect symmetry between the two limits: distinguishability (the lowest similarity) or nonsymmetry and indistinguishability (the highest similarity) or symmetry.

*2.2. Entropy and Information*

In statistical mechanics, entropy is a function of the distribution of the energy levels. The entropy concept in information theory is much more general. However, information theory, which has been used mainly in communication and computer science [10,11], is about the *process* of the communication channel. We will revise it to become a theory regarding the *structure* of a considered system. Process will be considered as a series of structures. Instead of defining *ensemble* (See: section 2.1, ref. [10]), we directly define that *the macroscopic structure* with regard to a certain kind of property $X$ of a considered system as a triple $(x, M_X, P_X)$ where the outcome $x$ is a *microscopic structure* (a microstate), which takes on one of a set of possible microstates, $M_X = \{m_1, m_2, ..., m_i, ..., m_w\}$, having probabilities $P_X = \{p_1, p_2, ..., p_i, ..., p_w\}$ with probability $P(x = m_i) = p_i$, $p_i \geq 0$ and $\sum_{m_i \in M_X} P(x = m_i) = 1$. The set $M$ is mnemonic for microscopic structures (or microstates). For a dynamic system, the structure is a mixture among the $w$ microstates (see examples illustrated in figures 3 and 4). In thermodynamics and statistical mechanics, the considered property is exclusively the energy level. In information theory, the property can be spin orientation itself (figure 3).

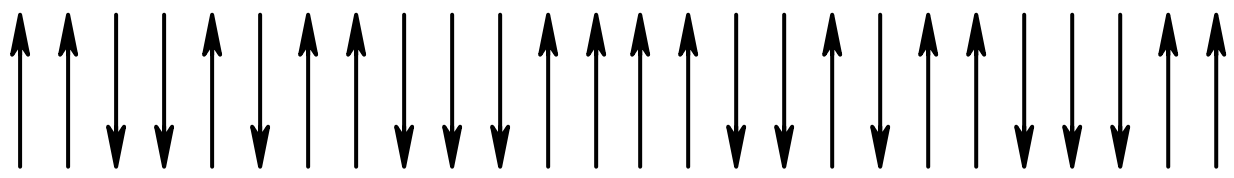

(a) A typical microstate as a freeze-frame of an array of spins undergoing up-and-down tumbling.

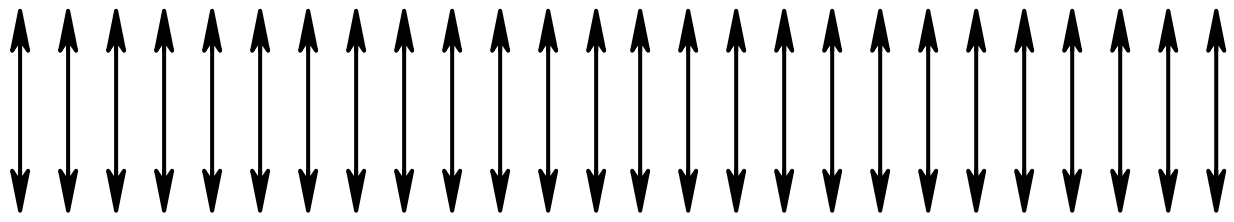

(b) The mixture of all $w$ microstates.

**Figure. 3.** Dynamic motion of spin-up and spin-down binary system. (a) A microstate. (b) The mixing of all microstates defines a symmetric macroscopic state.

ABACBC

(a) A microstate.

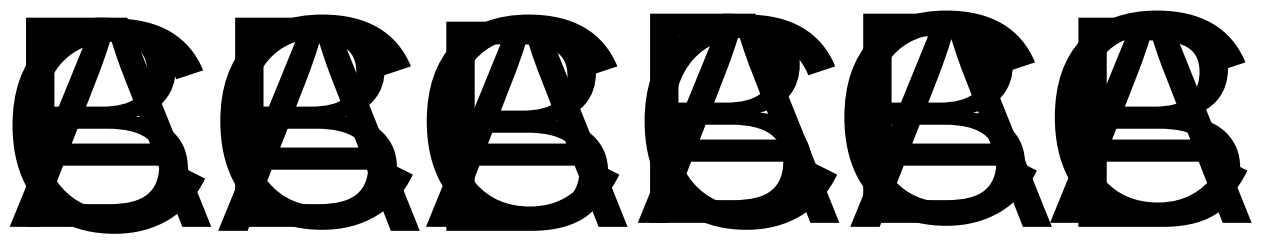

(b) The mixing of all 729 microstates.

**Figure 4.** (a) One of 729 ($3^N = 3^6 = 729$) microstates. (b) Schematic representation of information loss due to dynamic motion. The pictures at the six positions are the same, hence they give a symmetric macroscopic state. All the three letters appear at a position with the same probability. All the microstates also appear with the same probability.

Entropy $S$ of the structure regarding the property $X$ is given by the following familiar Gibbs-Shannon expression [11] (See also section 2.4, ref. [10]).

$$S = -\sum_{i=1}^{w} p_i \ln p_i \tag{1}$$

with the understanding that $0\ln 0 = 0$ and $1\ln 1 = 0$. Because $1 \ge p_i \ge 0$ and $\ln p_i < 0$, entropy is non-negative ($S \ge 0$). If the $w$ microstates have the *same* value of the considered property, hence the *same* value of $p_i$, $p_i = 1/w$ and

$$-\sum_{i=1}^{w} \frac{1}{w} \ln \frac{1}{w} = \ln w = L \tag{2}$$

The maximal entropy is denoted as $L$, because

$$\ln w \geq -\sum_{i=1}^{w} p_i \ln p_i \quad (3)$$

(This is the Gibbs inequality, see [7d] and the relevant citations). Entropy is a logarithmic function of $w_S$, the (apparent) symmetry number (or the apparent number of microstates of indistinguishable property, or the apparent number of the *equivalent* microstate which are of the *same* value of the considered property [7d]):

$$S = \ln w_S \quad (4)$$

Equation 4 is the familiar Boltzmann entropy expression. Combination of equations 1 and 4 leads to

$$w_S = \exp\left( -\sum_{i=1}^{w} p_i \ln p_i \right) \quad (5)$$

Obviously

$$w \geq w_S \quad (6)$$

*Example 2.* It can be illustrated by the entropy of a binary system (figure 4.1 in [10]) as a function of $p_i$. Coin tossing is a typical example of a binary system. Racemization of S and D enantiomers of $N$ molecules at high temperature, a ferromagnetic system (spin-up and spin-down) of $N$ spins at high temperature (figure 3) are examples in chemical physics. The number of microstates is $w = 2^N$. The maximum entropy is $\ln 2^N$ when $p_1 = p_2 = 0.5$ regarding the outcome of an interchanging enantiomer or a tumbling spin, or $p_1 = p_2 = ... = p_w = \frac{1}{2^N}$ for all the $w$ microstates. A microstate is a possible sequence of $N$ times of tossing or a possible picture of $N$ enantiomers or $N$ spins (figure 3).

Let us recall Lewis' remark that "gain of entropy means loss of information, nothing more ......" [12] and define the relationship of entropy and information. Lewis' remark also gives the hint that information can be converted to entropy. First, a new logarithmic function $L$ can be defined as the sum of entropy $S$ and information $I$:

$$L = S + I \quad (7)$$

$L$ is mnemonic for Logarithmic function or the "Largest possible value" of either the logarithmic function $S$ (equation 2) or $I$:

$$L = \ln w \quad (8)$$

Then, entropy is expressed as information loss [12]

$$S = L - I \quad (9)$$

or in certain cases when the absolute values are unknown,

$$\Delta S = \Delta L - \Delta I \quad (10)$$

for a change between two structures.

*Example 3.* For a 1.44 MB floppy disk, $L = 1.44$ MB whether the disk is empty or occupied with a file of the size of 1.44 MB. Let us use any available compression method to reduce the size of the original file to 0.40 MB. Then, $I = 0.40$ MB, $S = 1.04$ MB and $L = 1.44$ MB.

From equations 4 and 8, a logarithmic expression of information

$$I = \ln w_I \quad (11)$$

can be given, and

$$w_I = w / w_S \tag{12}$$

where $w_I$ is called the nonsymmetry number – the number of microstates of distinguishable property whereas $w_S$ in equation 4 called symmetry number (vide infra, section 6).

The logarithmic functions $S$, $I$ and $L$ are all nonnegative in value, dimensionless and they are all macroscopic properties.

From equation 9 and the fact that in practice information can be recorded only in a static structure, we may define static entropy [7d]. A macroscopic static structure is described by a set of $w$ microstates which are the $w$ possible rearrangements or $w$ possible transformations (recall the operations in group theory treatment of static structure which can be either an individual molecule or an assemblage) [2,8]).

In the entropy and information expressions, the unit is called nat (natural logarithmic unit). For a binary system, the unit is bit. Normally there is a positive constant in these expressions (e.g., the Boltzmann constant $k_B$ in thermodynamics). Here we put the constant as 1. In thermodynamics, we may denote the traditionally defined thermodynamics entropy as

$$S_T = k_B S \tag{13}$$

and

$$E = k_B TS + F \tag{14}$$

where $E$ is the total energy and $F$ the Helmholtz potential.

### *2.3. Labeling and Formatting*

In this paper similarity and its two limits (distinguishability and indistinguishability) will be frequently considered. Therefore, similarity and its significance should be clearly defined. Entropy or information (symmetry or nonsymmetry, vide infra) can be defined regarding a property $X$. Suppose there are $n$ kinds of property $X$, $Y$, $Z$, ..., etc. which are independent. In order to enumerate the number of possible microstates regarding $X$, particularly in the cases of symmetry study, where we may encounter the real problem of indistinguishability, labeling with some of the other $n$-1 kinds of property $Y$, $Z$, ..., etc. is necessary. Microscopically the components in the considered system may appear as several kinds of property $X$, $Y$, $Z$, ..., etc. If only one kind of property is detected by the instrument (e.g., a chemical sensor to detect the existence of certain gas molecule) or by a machine (e.g., a heat engine which detects the average kinetic energy of the gas molecules [6c]), the others can be used for labeling purposes to distinguish the otherwise indistinguishable individual components. We have the following postulate (or definition):

*Definition*: The distinguishability of the components (microstates, individual molecules or phases) due to the *labeling* does not influence the similarity or indistinguishability of the individuals or the existing distinguishability of the components (microstates, individual molecules or phases) regarding the considered property.

In other words, if a difference in a certain kind of property $Y$ does not influence the indistinguishability of the components regarding the considered $X$, the introduced or the existing difference in $Y$ is defined as labeling.

*Example 4*. For a coin tossing example (example 2), the precondition for $p_1 = p_2 = 0.5$ is that the relevant property of the two sides of the coin are the same (indistinguishable, or symmetric). The two different figures on the two sides of the coin can be regarded as two labels. Similarly, for enantiomers and spin orientations if the energy levels are the same, the chirality of the two different enantiomers $D$ and $L$ or the two different orientations can be used as two labels.

*Example 5*. Suppose information is recorded by a set of individual items, for example, the symbols 0 and 1 in a binary system used in computer science. The color or font size or the position index 1234567 does not influence the indistinguishability of a string of seven zeros: 0000000 or 0000000 or 1234567. Therefore these properties can be used for labeling.

*Example 6*. DNA is the genetic substance due to the distinguishability of the four organic bases (two pyrimidines – cytosine and thymine, and two purines – adenine and guanine). This kind of distinguishability does not influence the symmetric backbone structure (the periodicity of the sugar-phosphate backbone) along the axis that can be detected by X-ray diffraction [13]. Therefore the different organic bases can be used for labeling.

*Example 7*. The information retrieval problem. The distinguishable ID number or phone number belonging to an individual person does not influence the similarity of characters of the individuals or the indistinguishability of family names or given names.

*Example 8*. For an *ideal* gas model in thermodynamics, all the monatomic ideal gases are the same (e.g., He and Ar). All the diatomic ideal gases are also the same. The difference (e.g., nitrogen $N_2$ and oxygen $O_2$ gases, or the ordinary water $H_2O$ and deuterated water $D_2O$) can be regarded solely as labeling which does not influence the indistinguishability of the gases the heat engine experiences [6c].

Because the ideal gas model is very important, we summarize the conclusion made in the preceding example in the following theorem:

*Theorem: To a heat engine, all the ideal gases are indistinguishable.*

All ideal gases have the following state equation as detected in a heat engine:

$$PV = k_{\mathrm{B}} NT \tag{16}$$

where $P$ is the pressure and $T$ the temperature. The definition of labeling is very important for a final resolution of the Gibbs paradox of entropy of mixing (see section 8.1.2).

*Definition*: The indistinguishability of the components (microstates, individual molecules or phases, etc.) due to *formatting* does not influence the distinguishability of the individuals or the existing indistinguishability of the components (microstates, individual molecules or phases) regarding the considered property.

Normally formatting produces periodicity in a static structure. In DNA (example 6), the periodicity of the sugar-phosphate backbone can be regarded as formatting. In figure 3, the spin array is formatted as a string with an equal interval. In figure 4, the letters are formatted as the same font size and as a string with an equal interval. If the system is formatted with some of the other kinds of property *Y, Z, ...,* etc. the amount of information recorded by using the distinguishability of the property $X$ will not change regardless of whether we create symmetry regarding many other kinds of property. It will be soon made clear that formatting increases the stability of the system at the sacrifice of the information recording capacity. For a memory system like a hard disk in a computer, formatting is the necessary preparation for information recording to provide certain stability of the system.

### *3. The Three Laws and the Stability Criteria*

Parallel to the first and the second laws of thermodynamics, we have:

*The first law of information theory:* the logarithmic function $L$ ($L = \ln w$, or the sum of entropy and information, $L = S + I$ ) of an isolated system remains unchanged.

*The second law of information theory:* Information $I$ of an isolated system decreases to a minimum at equilibrium.

We prefer to use information *minimization* as the second law of information theory because there is a stability criterion of potential energy *minimization* in physics (see section 6). Another form of the second law is the same in form as that of the second law of thermodynamics: *for an isolated system, entropy S increases to a maximum at equilibrium*. For other systems (closed system or open system), we define (see: p. 623 of ref. [14])

$$\text{universe} = \text{system} + \text{surroundings} \tag{17}$$

and treat the universe formally as an isolated system (Actually we have always done this in thermodynamics following Clausius [9]). Then, these two laws are expressed as the following: *The function L of the universe is a constant. The entropy S of the universe tends toward a maximum.* Therefore, the second law of information theory can be used as the criteria of structural stability and process spontaneity (or process irreversibility) in all cases, whether they are isolated systems or not. If the entropy of system + surroundings increases from structure A to structure B, B is more stable than A. The higher the value $\Delta S$ for the final structure, the more spontaneous (or more irreversible) the process will be. For an isolated system the surroundings remain unchanged.

At this point we may recall the definition of equilibrium. Equilibrium means the state of indistinguishability (a symmetry, the highest similarity) [7d], e.g., as shown in figure 5, the thermal equilibrium between parts A and B means that their temperatures are the same (See also the so-called zeroth law of thermodynamics [14]).

We have two reasons to revise information theory. Firstly, the entropy concept has been confusing in information theory [15] which can be illustrated by von Neumann's private communication with Shannon regarding the terms entropy and information (Shannon told the following story behind the choice of the term "entropy" in his information theory: "My greatest concern was what to call it. I thought of calling it 'uncertainty'. When I discussed it with John von Neumann, he had a better idea: 'You should call it entropy, for two reasons. In the first place your uncertainty function has been used in statistical mechanics under that name, so it already has a name. In the second place, and more important, no one knows what entropy really is, so in a debate you will always have the advantage' " [15a]). Many authors use the two concepts entropy and information interchangeably (see also ref. 7d and citations therein). Therefore a meaningful discussion on the conversion between the logarithmic functions $S$ and $I$ is impossible according to the old information theory. Secondly, to the present author's knowledge, none of several versions of information theory has been applied to physics to characterize structural stability. For example, Jaynes' information theory [16] should have been readily applicable to chemical physics but only a parameter similar to temperature is discussed to deduce that at the highest possible value of that temperature-like parameter, entropy is the maximum. Jaynes' information theory or the so-called maximal entropy principle has been useful for statistics and data reduction (See the papers presented at the annual conference MaxEnt). Brillouin's negentropy concept [17] has also never been used to characterize structural stability.

To be complete, let us add the third law here also:

*The third law of information theory:* For a perfect crystal (at zero absolute thermodynamic temperature), the information is zero and the static entropy is at the maximum.

The third law of information theory is completely different from the third law of thermodynamics although the third law of thermodynamics is still valid regarding the dynamic entropy calculation (However, the third law of thermodynamics is useless for assessing the stabilities of different static structures). A more general form of the third law of information theory is "for a perfect symmetric *static* structure, the information is zero and the *static* entropy is the maximum". The third law of information theory defines the static entropy and summarizes the conclusion we made on the relation of static symmetry, static entropy and the stability of the static structure [7d] (The static entropy is independent of thermodynamic temperature. Because it increases with the decrease in the total energy, a negative temperature can be defined corresponding to the static entropy [7a]. The revised information theory suggests that a negative temperature also can be defined for a dynamic system, e.g., electronic motion in atoms and molecules [7a]).

The second law of thermodynamics might be regarded as a special case of the second law of information theory because thermodynamics treats only the energy levels as the considered property $X$ and only the dynamic aspects. In thermodynamics, the symmetry is the energy degeneracy [18,19]. We can consider spin orientations and molecular orientations or chirality as relevant properties which can be used to define a static entropy. What kinds of property besides the energy level and their similarities are relevant to structural stability? Are they the so-called observables [18] in physics? Should these properties be related to energy (i.e., the entropy is related to energy by temperature-like intensive parameters as differential equations where Jaynes' maximal entropy principle might be useful) so that a temperature (either positive or negative [7a]) can be defined? These problems should be addressed very carefully in future studies.

Similar to the laws of thermodynamics, the validity of the three laws of information theory may only be supported by experimental findings. It is worth reminding that the thermodynamic laws are actually postulates because they cannot be mathematically proved.

## 4. Similarity Principle and Its Proof

Traditionally symmetry is regarded as a discrete or a "yes-or -no" property. According to Gibbs, the properties are either the same (indistinguishability) or not the same (figure 1a). A continuous measure of static symmetry has been elegantly discussed by Avnir and coworkers [20]. Because the maximum similarity is indistinguishability (the sameness) and the minimum is distinguishability, corresponding to the symmetry and nonsymmetry, respectively, naturally similarity can be used as a continuous measure of symmetry (section 2.1). The similarity refers to the considered property $X$ of the components (see the definition of labeling in section 2.3) which affects the similarity of the probability values of all the $w$ microstates and eventually the value of entropy (equation 1).

Gibbs paradox statement [21] is a higher similarity–lower entropy relation (figure 1a), which has been accepted in almost all the standard textbooks of statistical mechanics and thermodynamics. The resolution of the Gibbs paradox of entropy–similarity relation has been very controversial. Some recent debates are listed in reference [21]. The von Neumann continuous relation of similarity and entropy (the higher the similarity among the components is, the lower value of entropy will be, according to his resolution of Gibbs paradox) is shown in figure 1b [21j]. Because neither Gibbs' discontinuous higher similarity–lower entropy relation (where symmetry is regarded as a discrete or a "yes-or -no" property ) nor von Neumann's continuous relation has been proved, they can be at most regarded as postulates or assumptions. Based on all the observed experimental facts [7d], we must abandon their postulates and accept the following postulate as the most plausible one (figure 1c):

*Similarity principle:* The higher the similarity among the components is, the higher the value of entropy will be and the higher the stability will be.

The components can be individual molecules, molecular moieties or phases. The similarity among the components determines the property similarity of the microstates in equation 1. Intuitively we may understand that, the values of the probability $p_i$ are related to each other and therefore depend solely on the similarity of the considered property $X$ among the microstates. If the values of the component property are more similar, the values of $p_i$ of the microstates are closer (equation 1 and 2). The similarity among the microstates is reflected in the similarity of the values of $p_i$ we may observe. There might be many different methods of similarity definition and calculation. However, entropy should be always a monotonically increasing function of any kind of similarity of the relevant property if that similarity is properly defined.

*Example 9.* Suppose a coin is placed in a container and shaken violently. After opening the container you may find that it is either head-up or head-down (The portrait on the coin is the label). For a bent coin, however, the property of head-up and the probability of head-down will be different. The entropy will be smaller and the symmetry (shape of the coin) will be reduced. In combination with example 2, it is easy to understand that entropy and similarity increase together.

Now, let us perform the following proof: the Gibbs inequality,

$$\ln w \geq -\sum_{i=1}^{w} p_i \ln p_i \tag{18}$$

has been proved in geometry (see the citations in [7d]). As $\ln w$ represents the maximum similarity among the considered $w$ microstates [18], the general expression of entropy $-\sum_{i=1}^{w} p_i \ln p_i$ must increase continuously with the increase in the property similarity among the $w$ microstates. The maximum value of $\ln w$ in equation 2 corresponds to the highest similarity. Finally, based on the second law of the revised information theory regarding structural stability that says that *the entropy of an isolated system (or system + environment) either remains unchanged or increases,* the similarity principle has been proved.

Following the convention of defining similarities [7d,7e] as an index in the range of [0,1], we may simply define

$$Z = \frac{S}{L} = \frac{\ln w_S}{\ln w} = \frac{-\sum_{i=1}^{w} p_i \ln p_i}{\ln w} \tag{19}$$

as a similarity index (figure 1c). As mentioned above, $Z$ can be properly defined in many different expressions and the relation that entropy increases continuously with the increase in the similarity will be still valid (However, the relation will not be necessarily a straight line if the similarity is defined in another way). For example, the $N^2$ similarities $r_{ij}$ in a table

$$\begin{matrix} r_{11} & r_{12} & \cdots & \cdots & r_{1N} \\ r_{21} & r_{22} & \cdots & \cdots & r_{2N} \\ \cdots & \cdots & \cdots & \cdots & \cdots \\ \cdots & \cdots & \cdots & \cdots & \cdots \\ r_{N1} & r_{N2} & \cdots & \cdots & r_{NN} \end{matrix} \tag{20}$$

can be used to define a similarity value of a system of $N$ kinds of molecule [7c,7e].

## 5. Curie-Rosen Symmetry Principle and Its Proof

It is straightforward to prove the higher symmetry–higher stability relation (the Curie-Rosen symmetry principle [2]) as a special case of the similarity principle. Because higher similarity is correlated with a higher degree of symmetry, the similarity principle also implies that entropy can be used to measure the degree of symmetry. We can conclude: *The higher the symmetry (indistinguishability) of the structure is, the higher the value of entropy will be.* From the second law of information theory, the higher symmetry–higher entropy relation is proved.

The proof of the higher symmetry–higher entropy relationship can be performed by contradiction also. Higher symmetry number–lower entropy value relation can be found in many textbooks (see citations in [7d] and p. 596 of [19]) where the existence of a symmetry would result in a decrease in the entropy value:

$$\Delta S = -\ln s \tag{21}$$

where σ denotes the symmetry number and $s \geq 1$. Let the entropy change from $S'$ to $S$ is

$$\Delta S = S' - S = -\ln s \tag{22}$$

Then the change in symmetry number would be a factor

$$s = \frac{w_{S'}}{w_S}, \tag{23}$$

and

$$S' - S = -\ln \frac{w_{S'}}{w_S} = -\ln w_{S'} - (-\ln w_S) \tag{24}$$

where $w_S$ and $w_{S'}$ are the two symmetry numbers. This leads to an entropy expression $S = -\ln w_S$. However, because any structure would have a symmetry number $w_S \geq 1$, entropy would be a negative value, which contradicts the definition that entropy is always positive. Therefore neither $\Delta S = -\ln s$ nor $S = -\ln w_S$ are valid. The correct form should be equation 4 ($S = \ln w_S$, $w_S \geq 1$). In combination with the second law of information theory, the higher symmetry–higher stability relation (the symmetry principle) is also proved.

Rosen discussed several forms of symmetry principle [2]. Curie's causality form of the symmetry principle is that *the effects are more symmetric than the causes*. The higher symmetry–higher stability relation has been clearly expressed by Rosen [2]:

*For an isolated system the degree of symmetry cannot decrease as the system evolves, but either remains constant or increases.*

This form of symmetry principle is most relevant in form to the second law of thermodynamics. Therefore the second law of information theory might be expressed thus: an isolated system will evolve spontaneously (or irreversibly) towards the most stable structure, which has the highest symmetry. For closed and open systems, the isolated system is replaced by the universe or system + environment.

Because entropy defined in the revised information theory is more broad, symmetry can include both static (the third law of information theory) and dynamic symmetries. Therefore, we can predict the symmetry evolution from a fluid system to a static system when temperature is gradually reduced: the most symmetric static structure will be preferred (vide infra).

## 6. A Comparison: Information Minimization and Potential Energy Minimization

A complete structural characterization should require the evaluation of both the degree of symmetry (or indistinguishability) and the degree of nonsymmetry (or distinguishability). Based on the above discussion we can define entropy as the logarithmic function of symmetry number $w_S$ in equation 4. Similarly, the number $w_I$ can be called nonsymmetry number in equation 11.

Other authors briefly discussed the higher symmetry–higher entropy relation previously (see the citation in [2]). Rosen suggested that a possible scheme for symmetry quantification is to take for the degree of symmetry of a system the order of its symmetry group (or its logarithm) (p.87, reference [2a]). The degree of symmetry of a considered system can be considered in the following increasingly simplified manner: in the language of group theory, the group corresponding to $L$ is a direct product of the two groups corresponding to the values of entropy $S$ and information $I$:

$$\mathrm{G} = \mathrm{G}_S \times \mathrm{G}_I \tag{25}$$

where $\mathrm{G}_S$ is the group representing the observed symmetry of the system, $\mathrm{G}_I$ the nonsymmetric (distinguishable) part that potentially can become symmetric according to the second law of information, and $\mathrm{G}$ is the group representing the maximum symmetry. In the language of the numbers of microstates

$$w = w_S \cdot w_I \tag{26}$$

where the three numbers are called the maximum symmetry number, the symmetry number and the nonsymmetry numbers, respectively. These numbers of microstate could be the orders of the three groups if they are finite order symmetry groups. However, there are many groups of infinite order such as those of rotational symmetry, which should be considered in detail in our further studies.

The behavior of the logarithmic functions $L = \ln w$, $S = \ln w_S$ and their relation

$$L = S + I \tag{7}$$

can be compared with that of the total energy $E$, kinetic energy $E_K$ and potential energy $E_P$ which are the eigenvalues of $H$ and $K$ and $P$ in the Hamiltonian expressed conventionally in either classical mechanics or quantum mechanics as two parts:

$$H = K + P \tag{27}$$

The energy conservation law and the energy minimization law regarding a spontaneous (or irreversible) process in physics (In thermodynamics they are the first and the second law of thermodynamics) says that

$$\Delta E = \Delta E_K + \Delta E_P \tag{28}$$

and $\Delta E = 0$ (e.g., for a linear harmonic oscillator, the sum of the potential energy and kinetic energy remains unchanged), $\Delta E_P \leq 0$ for an isolated system (or system +environment). In thermodynamics, Gibbs free energy $G$ or Helmholtz potential $F$ (equation 14) are such kinds of potential energy. It is well known that the minimization of the Helmholtz potential or Gibbs free energy is an alternative expression of the second law of thermodynamics. Similarly,

$$\Delta L = \Delta S + \Delta I \tag{29}$$

For a spontaneous process of an isolated system, $\Delta L = 0$ (the first law of information theory) and $\Delta I \leq 0$ (the second law of information theory or the minimization of the degree of nonsymmetry) or $\Delta S \geq 0$ (the maximization of the degree of symmetry). For an isolated system,

$$\Delta S = -\Delta I \tag{30}$$

For systems that cannot be isolated, spontaneous processes with both $\Delta S > 0$ or $\Delta S < 0$ for the systems are possible provided that

$$\Delta S > 0 \tag{31}$$

for the universe ($\Delta S = 0$ for a reversible process. $\Delta S < 0$ are impossible process). The maximum symmetry number, the symmetry number and the nonsymmetry number can be calculated as the exponential of the maximum entropy, entropy and information, respectively. These relations will be illustrated by some examples and will be studied in more detail in the future.

The revised information theory provides a new approach to understanding the nature of energy (or energy-matter) conservation and conversion. For example the available potential energy due to distinguishability or nonsymmetry can be calculated. This can be illustrated in a system undergoing spontaneous mass or heat transfer between two parts of a chamber (figure 5). The distinguishability or nonsymmetry is the cause of the phenomena. Many processes are irreversible because a process does not proceed from a symmetric structure to a nonsymmetric structure.

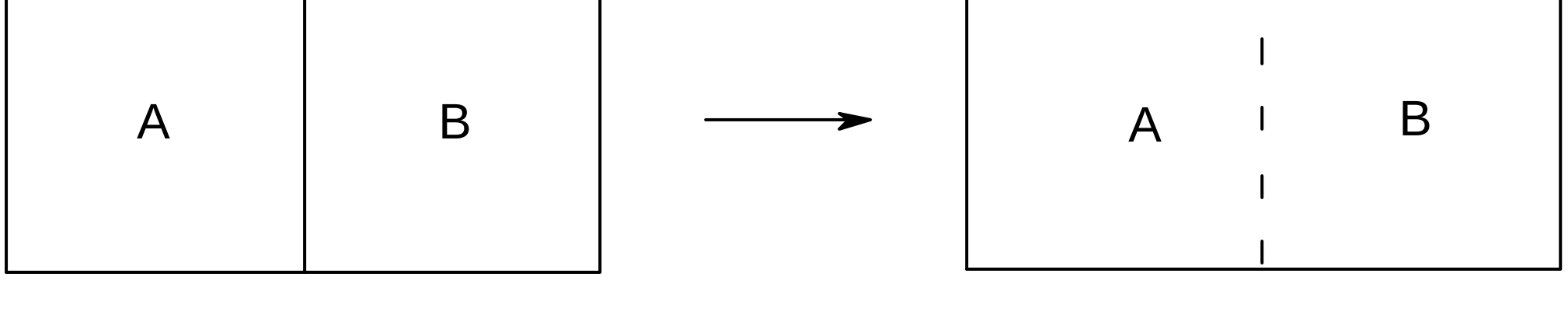

**Figure 5.** Mass (ideal gas) transfer and heat transfer between two parts of a chamber after removal of the barrier. The whole system is an isolated system.

## 7. Interactions

From the second law of information theory, any system with interactions among its components will evolve towards an equilibrium with the minimum information (or maximum symmetry). This general principle leads to the criteria of equilibrium for many kinds of system. For example, in a mechanical system the equilibrium is the balance of forces and the balance of torques.

We may consider symmetry evolution generally in two steps. Step one: bring several components (e.g., parts A and B in figure 5) to the vicinity as individually isolated parts. We may treat this step as if these parts have no interaction. Step two: let these parts interact by removing electromagnetic insulation, thermal insulation, or mechanical barrier, etc. (e.g., the right side of figure 5).

For step one, similarity analysis will be satisfactory. For the example shown in figure 5, we measure if the two parts A and B are of the same temperature (if they are, there will be no heat transfer), the same pressure (if they are, there will be no mass transfer) or the same substances (if they are, there will be no chemical reactions.

We may calculate the total logarithmic function $L_{\text{total}}$, total entropy ($S_{\text{total}}$) and total information of an isolated system of many parts for the first step. In the same way, we may calculate a system of complicated, hierarchical structures. Suppose there are $M$ hierarchical levels ($i = 1,2,..., M$ , e.g., one level is a galaxy, the other levels are the solar system, a planet, a box of gas, a molecule, an atom, electronic motion and nuclear motion inside the atom, subatomic structures, etc.) and $N$ parts ($j = 1,2,..., N$ , e.g., different cells in a crystals, individual molecules, atoms or electrons, or spatially different locations). It should be greater than the sum of the individual parts at all the hierachical levels,

$$S_{\text{total}} \geq \sum_{i,j} S_{i,j} \tag{32}$$

because any assembling and any interaction (coupling, etc.) among the parts or any interaction between the different hierachical levels will lead to information loss and entropy increase. Many terms of entropy due to interaction should be included in the entropy expression.

*Example 10.* The assembling of the same molecules to form a stable crystal structure is a process similar to adding many copies of the same book to the shelves of one library. Suppose there are 1GB information in the book. Even though there are 1,000,000 copies, the information is still the same and the information in this library is $I_{\text{library}} = \sum_{j=1}^{1000000} I_j = I_1 = 1\text{ GB}$. The entropy is 999999 GB.

*Example 11.* Interaction of the two complementary strands of polymer to form DNA. The combination leads to a more stable structure. The system of these two components (part 1 and part 2) have entropy greater than the sum of the individual parts ($S_{1+2} \geq 2S_1$) because $I_{1+2} = \sum_{j=1}^{2} I_j = I_1 \leq 2I_1$.

However, operator theory or functional analysis might be applied for more vigorous mathematical treatment, which will be presented elsewhere. Application of the revised information theory to specific cases with quantitative calculations also will be topics of further investigations. We will only outline some applications of the theory in the following sections.

Finally, we claim that due to interactions, the universe evolves towards maximum symmetry or minimum information.

## 8. Dynamic and Static Symmetries

There are two types of symmetries: *dynamic* symmetry and *static* symmetry. Both are related to information loss as schematically illustrated in figures 6 and 7. Dynamic entropy is the logarithm of the number of the microstates of identical property (or identical energy level in thermodynamics).

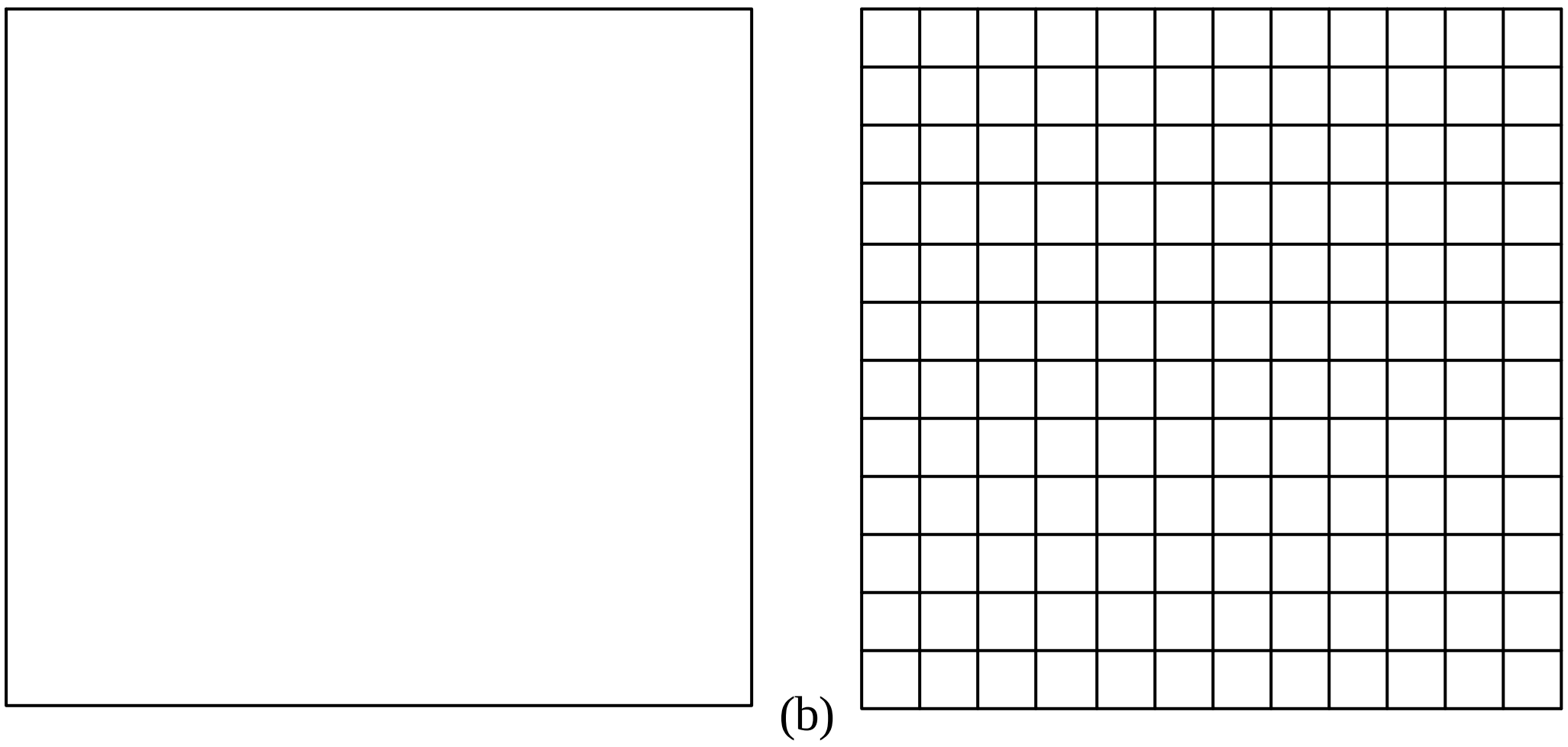

(a) (b)

**Figure 6.** Schematic representation of a highly symmetric system of an ideal gas (dynamic symmetry, very homogenous and isotropic) (a) or a perfect crystal (static symmetry) (b) and the information loss. The highly symmetric paintings like (a) and (b) found in many famous modern art museums are the emperor's new clothes. These paintings are “reproduced” here without courtesy.

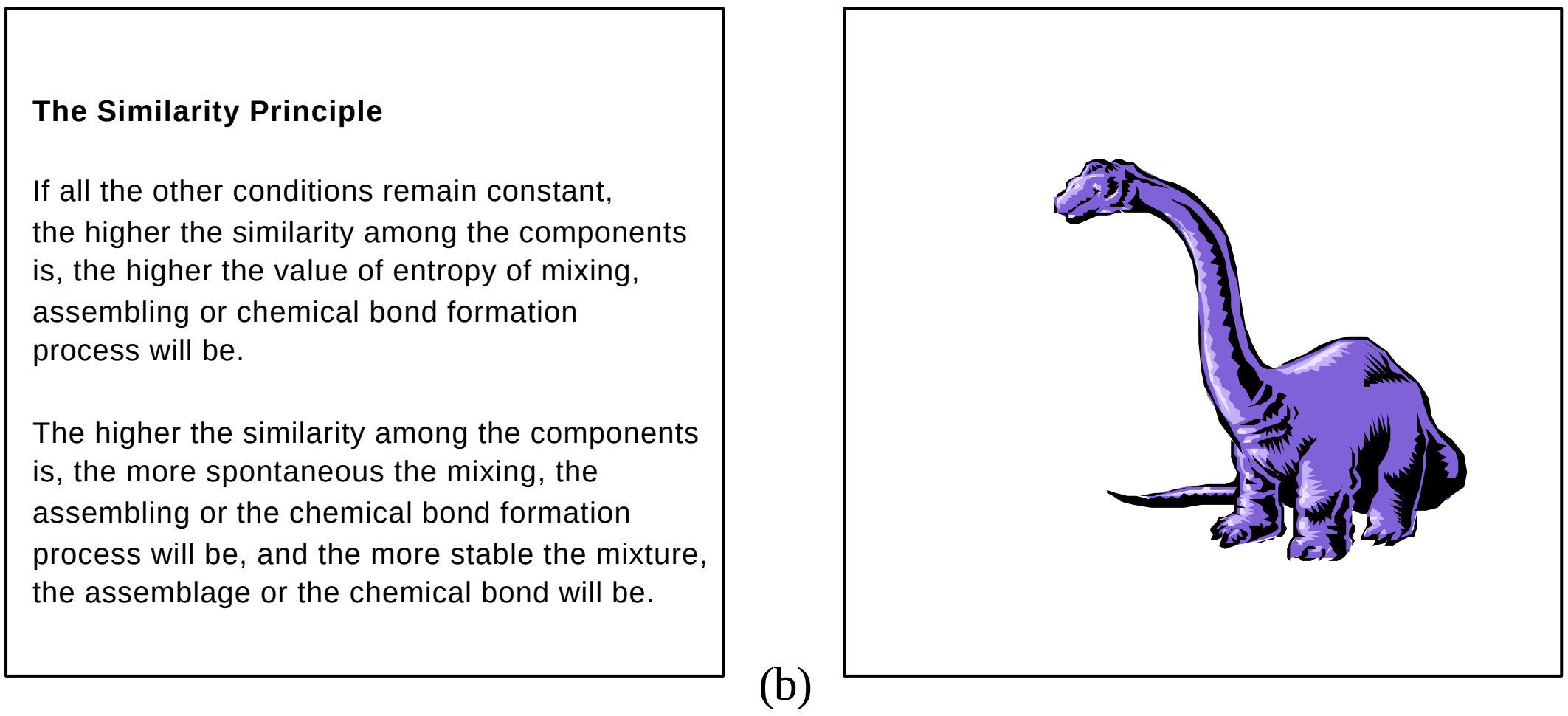

(b)

**Figure 7.** A static structure with certain text (a) and graphic (b) information recorded must have dramatically reduced symmetry.

The example of dynamic symmetry is the homogeneity of fluid systems of many particles, such as a gas of many molecules. The homogeneity of an ideal gas is due to the dynamic symmetry of the fluid phase. The static symmetry number can be easily estimated for a crystal [7d].

### *8.1. Dynamic Symmetry*

#### 8.1.1. Fluid Systems

Because information is recorded in static structures, systems of molecular dynamic motion have zero value of information and, therefore, the logarithmic functions $L$ and $S$ have equal value at the equilibrium state. In order to accommodate the kinetic energy of the system, the structure of the system cannot stay in only one of many accessible microstates. The macroscopic properties – entropy and symmetry (homogeneity, isotropicity) of an idea gas used in a heat engine can be considered first because the ideal gas model is most important in thermodynamics.

*Example 12.* Ideal gas mass transfer and heat transfer between two parts. Entropy of an ideal gas in thermodynamics is a special kind of dynamic entropy. For a free expansion process of an ideal gas within an isolated system with two parts of different pressure (figure 5), $\Delta E = 0$, $\Delta G < 0$, $\Delta I < 0$, $\Delta S > 0$ and $\Delta L = 0$ in thermodynamics and in our revised information theory treatment. Symmetry is higher at the final equilibrium structure: both sides are of the same pressure and temperature at the final structure.

8.1.2. Gibbs Paradox

In this context let us resolve Gibbs paradox of entropy of mixing [21]. It says that the entropy of mixing *decreases discontinuousl*y with an increase in similarity. It has a zero value for mixing of the indistinguishable subsystems (figure 1a). The isobaric and isothermal mixing of one mole of an ideal fluid A and one mole of a different ideal fluid B (figures 5 and 8) has the entropy increment

$$(\Delta S)_{\text{distinguishable}} = 2R\ln 2 = 11.53 \text{ J K}^{-1} \tag{33}$$

where $R$ is the gas constant, while

$$(\Delta S)_{\text{indistinguishable}} = 0 \tag{34}$$

for the mixing of indistinguishable fluids [21]. It is assumed that the two equations (33) and 33) are also applicable to the formation of solid mixtures and liquid mixtures (citations in [7d]). Gibbs paradox statement of entropy of mixing has been regarded as the theoretical foundation of statistical mechanics [23], quantum theory [24] and biophysics [21e]. It is certainly a most important problem in information theory if one intends to apply information theory to physics: e.g., Jaynes, the father of the maximal entropy principle [16], also considered this problem [21h]. The resolutions of this paradox have been very controversial for over one hundred years. Many famous physicists confidently claim that they have resolved this paradox, in very diverse and different ways. A sustaining problem is that they do not agree with one another. For example, besides numerous other resolutions [21], von Neumann [21j] provided a well-known quantum mechanical resolution of this paradox with which not many others are in full agreement [21d].

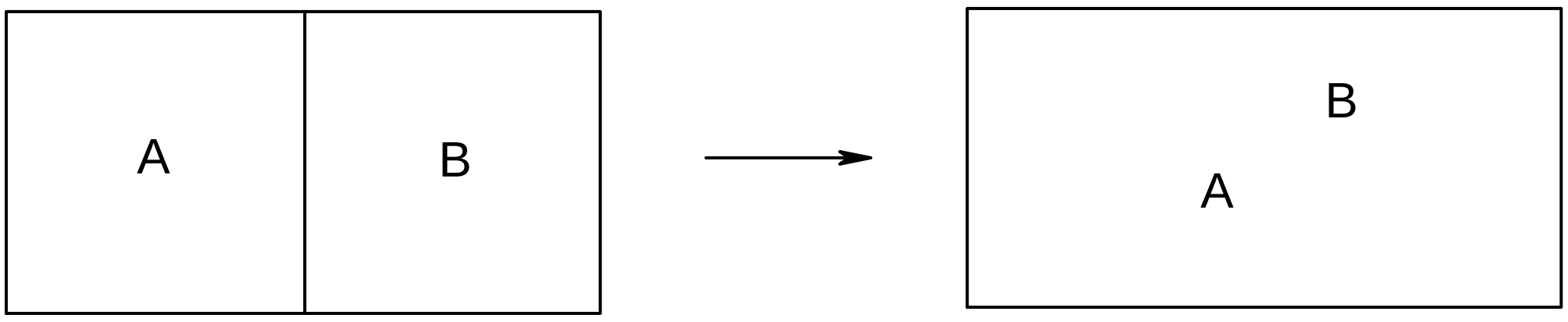

**Figure 8.** Ideal gas mixing in a rigid chamber.

Gibbs paradox is a problem regarding the relation of indistinguishability (symmetry) and entropy. It can be easily resolved if we recall the definition of labeling (section 2.3). For a rigid container, mixing two ideal gases is an identical process whether they are indistinguishable or distinguishable ideal gases, provided that the two gas chambers are parts of a rigid container (Because deformation changes the shape symmetry and entropy [7a], we suppose that the container is rigid. The interesting topic of shape symmetry evolution or deformation [7a] will be discussed in detail elsewhere): $\Delta E = 0$, $\Delta G = 0$, $\Delta(k_B TS) = 0$, and $\Delta S = 0$. As has been actually shown in experiment there is no change in the total energy, in the Gibbs free energy and there is no heat effect for the isobaric, isothermal mixing process. Therefore, the entropy change must be zero in both cases whether it is a mixing of the same ideal gases or of different ideal gases (equation 35):

$$(\Delta S)_{\text{distinguishable}} = 0 \tag{35}$$

8.1.3. Local Dynamic Motion in Solids (Crystals)

Another excellent example of local dynamic motion in a static structure is Pauling's assessment of residual entropy of ice [7d, 24]. The spin-up and spin-down binary system is shown in figure 3. The information loss and symmetry increase due to local dynamic motion can be further illustrated by typewriting different fonts at one location, as shown in figure 4. In many cases, the formation of certain static periodic structures can be regarded as formatting (section 2.3).

### *8.2. Static Symmetry*

If the temperature is gradually reduced, a system of many molecules may become a static structure. There are many possible static structures. In principle, for a condensed phase, the static structure can be any of the $w$ microstates accessible before the phase transition. Symmetric static structure (crystal) and nonsymmetric static structure must have different information (figures 6 and 7). Our theory predicts that the most symmetric microstate will be taken as the most stable static structure (the third law of information theory and the symmetry principle). Before the phase transition at higher temperature, the system should be of the highest possible dynamic symmetry (figure 9, for example). After the phase transition to form a solid phase, the system should evolve to the highest possible static symmetry (figure 10).

$$\begin{matrix} \leftrightarrow & \leftrightarrow & \leftrightarrow & \leftrightarrow & \leftrightarrow & \leftrightarrow & \leftrightarrow \\ \leftrightarrow & \leftrightarrow & \leftrightarrow & \leftrightarrow & \leftrightarrow & \leftrightarrow & \leftrightarrow \\ \leftrightarrow & \leftrightarrow & \leftrightarrow & \leftrightarrow & \leftrightarrow & \leftrightarrow & \leftrightarrow \\ \leftrightarrow & \leftrightarrow & \leftrightarrow & \leftrightarrow & \leftrightarrow & \leftrightarrow & \leftrightarrow \end{matrix}$$

**Figure 9.** A two-dimensional static array with local dynamic symmetry. At a position, the two orientations have identical probability. The highest local dynamic symmetry leads to the least information. The information loss is equivalent to printing many pages of a book on one page which will lead to symmetry (every position will be the same hybrid of the two orientations) and information loss.

$$\begin{matrix} \rightarrow & \rightarrow & \rightarrow & \rightarrow & \rightarrow & \rightarrow & \rightarrow \\ \rightarrow & \rightarrow & \rightarrow & \rightarrow & \rightarrow & \rightarrow & \rightarrow \\ \rightarrow & \rightarrow & \rightarrow & \rightarrow & \rightarrow & \rightarrow & \rightarrow \\ \rightarrow & \rightarrow & \rightarrow & \rightarrow & \rightarrow & \rightarrow & \rightarrow \end{matrix}$$

**Figure 10.** A schematic representation of a two-dimensional static system of high static symmetry with only one spin orientation.

Actually these examples are discussed in the famous Ising model (part III of the book [15a]) and the related phase transition problem [6]. The structure can be easily characterized by the spin orientation indistinguishability (a symmetry). Figure 3 is a one-dimensional array of spins. A two-dimensional system of spins at fixed lattices are given in figures 9 (dynamic symmetry) and 10 (static symmetry).

The expression of the entropy of mixing has been applied in the same way to the mixing processes to form gaseous, fluid or solid mixtures (see any statistical mechanics textbook). Therefore, in the Ising model, the entropy of a noncrystal structure would have a higher entropy due to the entropy of mixing *different* species according to the traditional expression (figure 1a). However, the most symmetric static structure (figure 10, for instance) has the highest static entropy according to our third law. The present author believes that the traditional way of calculating the entropy of mixing is the largest flaw in solid state physics in particular and in physics in general. Prigogine's dissipative structure theory (which has been claimed to have solved such kind of symmetry breaking problems [4, 25]), Wilson's method of renormalization group [5,26] and many other theories have been proposed. The symmetry breaking problem remains to be solved. According to our theory, the static symmetry (and the static entropy) should dominantly contribute to the stability of a ferromagnetic system in the spin-parallel static state below the Curie temperature. Due to the static symmetry, a perfect crystal has the highest static entropy (the third law of information theory). This conforms to perfectly the observed highest stability of crystal structure among all the possible static (solid) structures. The perfect crystal structure is equivalent to a newly formatted disk (hard disk or floppy disk) which has the highest symmetry and the least (or zero) information (figure 2b). Noncrystal solid structures are less stable (This prediction has already been confirmed by an abundance of experimental observations [7]; ask any experimental chemists or material scientists!).

## 9. Phase Separation and Phase Transition (Symmetry Breaking)

When the thermodynamic temperature decreases, there will be a phase separation where different substances separate as a result of the spontaneous assembling of the indistinguishable substances.

### *9.1. Similarity and Temperature*

Let us illustrate the relation of similarity in thermodynamics and the thermodynamic temperature $T$ with the simplest example. The energy level similarity between two energy levels $E_a$ and $E_b$ is calculated from their Boltzmann factors $e^{-\frac{E_a}{kT}}$ and $e^{-\frac{E_b}{kT}}$. The similarity will approach the maximum if temperature increases and the minimum if the temperature approaches zero:

$$\lim_{T\to\infty} e^{-\left|\frac{\Delta E}{kT}\right|} = \lim_{T\to\infty} e^{-\left|\frac{E_a - E_b}{kT}\right|} = 1 \tag{36}$$

$$\lim_{T\to 0} e^{-\left|\frac{\Delta E}{kT}\right|} = \lim_{T\to 0} e^{-\left|\frac{E_a - E_b}{kT}\right|} = 0 \tag{37}$$

Jaynes' maximal entropy principle [15a,16] may be applied to discuss the relation of similarity and a temperature-like parameter. Generally speaking, similarity increases with the increase in the absolute value of temperature $|T|$ (For a system of negative temperature where $T \le 0$, similarity increases with the increase in $|T|$; see ref. 7a).

*9.2. Phase Separation, Condensation, and the Densest Packing*

The similarity among components will decrease at a reduced temperature (equation 37). Because a heterogeneous structure with components of very different properties has high information and is unstable, phase separation for a system of multiple components will follow the similarity principle: *different components spontaneously separate. The components of the same (or very similar) properties mix or merge to form a homogeneous phase. The different components separate as a consequence of the assembling of components of the most similar (or the same) properties.* Spontaneous phase separation and its information loss (or symmetry increase) as well as the opposite process can be illustrated in figure 11 [7d]. Therefore, if phase separation is desirable, we decrease the temperature (equation 37); if the mixing of components is required, we increase the temperature (equation 36).

Spontaneous phase separation
or self-aggregation
or self-organization

$$10101000011101 \rightleftharpoons 00000001111111$$

Information registration

**Figure 11.** Informational registration as a binary string and the information loss due to a possible spontaneous process.

Spontaneous phase separation has been the main chemical process to separate different substances and to purify substances. For example, even enantiomers can be separated in this way [7d].

Condensation from a gaseous phase and many examples of the densest molecular packing in the formation of crystal structures and other solid state structure can be explained by our revised information theory as a phase separation of a binary system of two "species": substance and vacuum. In this approximation, all molecules (not necessarily the same kind of molecule or one kind of molecule) can be taken as the species "1" and all the parts of free space are taken as species "0". At a reduced thermodynamic temperature, substance and vacuum both occupy the space, but otherwise are so different that they will separate as a consequence of assembling the same species together as two "phases", one is the condensed substance phase, the other is a bulky "phase" of vacuum (figure 11).

*9.3. Phase transition and Evolution of the Universe and Evolution of Life*

The evolution of the Universe is a series of phase separations and phase transition. At extremely high temperature, even matter and antimatter can apparently coexist. At high temperature, the similarity is high (equation 36). When the temperature is reduced, the increase of the entropy can be achieved either by phase separation or phase transition as spontaneous processes. At reduced temperature, matter aggregates with matter. Then a mixture of matter and antimatter are not stable because their similarity is extremely low. The biosphere has only L-amino acids and D-sugars [27]. Our theory can shed light on the solution of the symmetry breaking phenomena during the universe evolution in general and the molecular evolution of life in particular. It is clear that at every stage of symmetry breaking, the critical phenomenon is characterized by the system's tendency towards the highest possible dynamic symmetry at higher temperature and the highest possible static symmetry at lower temperature.

## 10. Similarity Rule and Complementarity Rule

Generally speaking, *all* intermolecular processes (molecular recognition and molecular assembling or the formation of any kinds of chemical bond) and intramolecular processes (protein folding [28], etc.) between molecular moieties are governed either by the similarity rule or by the complementarity rule or both.

Similarity rule (a component in a molecular recognition process loves others of like properties, such as hydrophobic interaction, $\pi$-stacking in DNA molecules, similarity in softness of the well-known hard-soft-acid-base rules) predicts the affinity of individuals of *similar* properties. On the contrary, complementarity rule predicts the affinity of individuals of certain *different* properties. Both types of rule still remain strictly empirical. The similarity rule can be given a theoretical foundation by the similarity principle (figure 1c) [7a] after rejection of Gibbs' (figure 1a) and revised (figure 1b) relations of entropy–similarity.

All kinds of donor–acceptor interaction, such as enzyme and substrate combination, which may involve hydrogen bond, electrostatic interaction and stereochemical key-and-lock docking [30] (e.g., template and the imprinted molecular cavity [6b]), follow the complementarity rule. For the significance of the complementarity concept in chemistry, see the chapter on Pauling in the book [1b].

*Definition:* Suppose there are $n$ kinds of property $X$, $Y$, $Z$, ..., etc. (See the definition of entropy and labeling in section 2) and $n = l + m$. For a binary system, if the two individuals contrast in $l$ kinds of property (negative charge-positive charge or convex and concave, etc.) and exactly the same for the rest $m$ kinds of property, the relation of these two components is complementary. An example is given in figure 12.

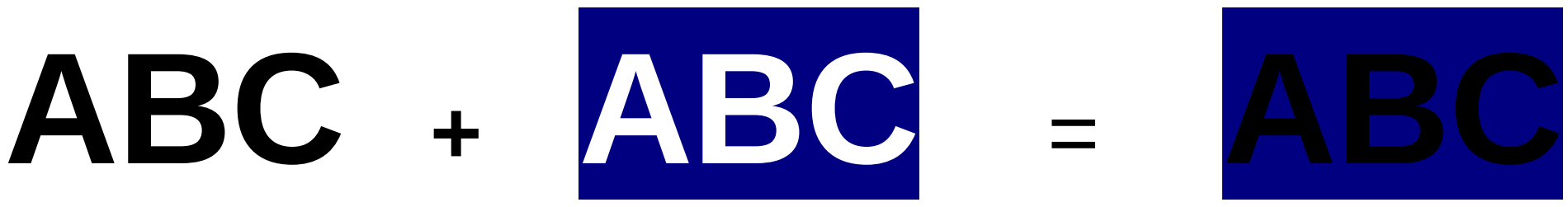

**Figure 12.** The print and the imprint [6b] are complementary.

Firstly, for stereochemical interaction (key-and-lock docking) following a complementarity rule it can be treated as a special kind of phase separation where the substance and the vacuum (as a different species) separate (cf. section 8). The final structure is more "complete", more integral, more "solid" and more symmetric. More generally speaking, the components in the structure of the final state become more similar due to the property *offset* of the components and the structure is more symmetric. The calculation of symmetry number, entropy and information changes during molecular interaction are numerous topics for further studies.

*Complementarity principle:* The final structure is more symmetric due to the property offset of the components. It will be more stable.

In the recent book by Hargittai and Hargittai [1b] many observations showed that in our daily life symmetry created by combination of parts means beauty. We may speculate that, in all these cases, the stability of the interaction of the symmetric static images (crystal, etc.) and symmetric dynamic processes (periodicity in time, etc.) may play an important role in our perception. Perception of visual beauty might be our visual organ's interaction with the images or a sequences of images and a tight interaction might result if the similarity rule and the complementarity rule are satisfied.

Normally we consider complementarity of a binary system (two partners). However, it can be an interaction among many components. The component motifs are *distinguishable* (for a binary system they are *contrast*) in the considered property or properties.

Chemical bond formation and all kinds of other interaction following the similarity and complementarity rules will lead to more stable *static* structures. Actually there are also similarity and complementarity rules for the formation of a stable *dynamic* system. The resonance theory is an empirical rule [29]. Von Neumann tried very hard to use the argument of entropy of mixing but his final conclusion regarding entropy–similarity relation (figure 1b) was wrong and cannot be employed to explain the stability of quantum systems [21j]). Pauling's resonance theory can be supported by the two principles – similarity principle and complementarity principle for a dynamic system regarding electron motion and electronic configuration [7a]. The final structure as a hybrid or a mixture of a "complete" set (having all the microstates or all the available canonical structures of very similar energy levels, similar configuration, etc. see 1 and 2 in figure 13a) should be more symmetric. The most complete set of microstates will lead to the highest dynamic entropy and the highest stability. Other structures (e.g., 3 in figure 13a) cannot contribute significantly to the structure because they are very different from the structure of 1 or 2.

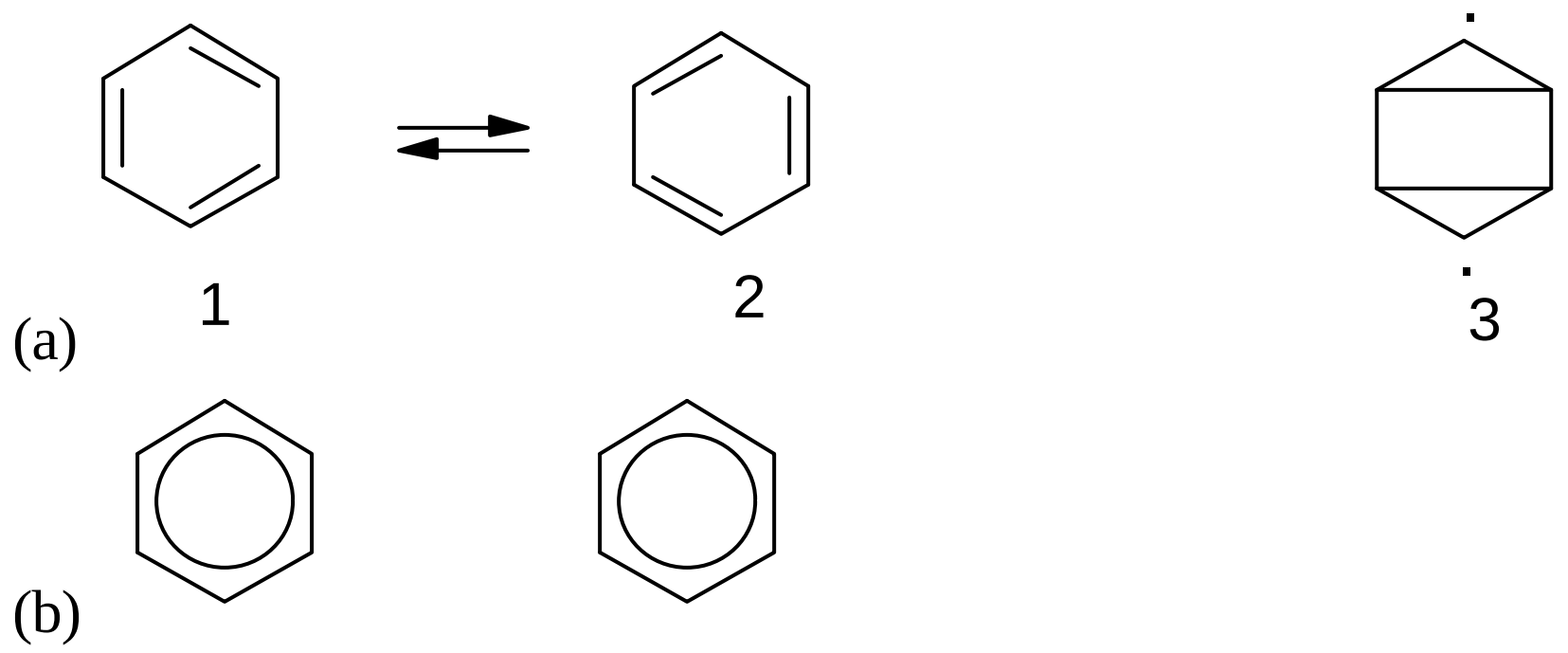

**Figure 13.** Information loss due to dynamic electronic motion. The different orientations of the valence bond benzene structures 1 and 2 in (a) can be used for recording information. However, the oscillation makes all the individual benzene molecules the same as shown in (b).

To explain the stability of the systems other than molecular systems, one may replace molecules by individuals or components under consideration and then calculate in a similar way as for molecular systems. At least the relative structural stability may be assessed for different structures where the similarity principle is applicable.

Finally we should emphasize that the similarity rule is always more significant than the complementarity rule, because most properties of the complementary components are the same or very similar ($m>l$). Examples in chemistry are the HSAB (hard-soft-acid-base) rule where the two components (acid and base) should have similar softness or similar hardness (see any modern texts in inorganic chemistry). Another example is the complementary pair of LUMO (lowest unoccupied molecular orbital) and HOMO (highest occupied MO) where the energy levels of the MO are very close (see a modern textbook of organic chemistry). The formation of a chemical bond can be illustrated by a stable marriage (example 13).

*Example 13*. Successful marriage might be a consequence of the highest possible similarity and complementarity between the couple. The stability of the relation is directly determined by how identical the partners are: both man and woman have interests in common, have the same goals, common beliefs and principles, and share in wholesome activities. Love and respect between the two are the complementarity at work which is the more interesting aspect of the marriage due to their contrasting attributes: One is submissive, the other not. A beautiful marriage is determined by how adaptable both of them are regarding differences, rather than by how identical they are. God created man and woman to make one as a complement of the other (Genesis 2:18).

## 11. Periodicity (Repetition) in Space and Time

The beauty of periodicity (repetition) or translational symmetry of molecular packing in crystals must be attributed to the corresponding stability. Formatting of a floppy disk or a hard disk will create symmetry and erasure of all the information. Formatting is a necessary preparation for stable storage of information where certain kinds of symmetry are maintained: e.g., books must be "formatted" by pages and lines where periodicity is maintained as background. The DNA molecules are "formatted" by the sugar and phosphate backbone as periodic units.

Steady-state is a special case of a dynamic system. Its stability depends on the symmetry (periodicity in time). The long period of chemical oscillation [1b,4] happens in a steady-state system. Cars on a highway must run in the same direction at the same velocity. Otherwise, if one car goes much slower, there will be a traffic accident (Curie said “nonsymmetry leads to phenomena” [2]), and the unfortunate phenomenon is the collision. Many kinds of cycle (heart-beating cycle, sleep-and-wake-up cycles, etc.) are the periodicity in time, which contributes stability, very much the same as the Carnot cycle of an ideal heat engine. However, exact repetition of everyday life must be very boring, even though it makes life simple or easy and stable (vida infra).

## 12. Further Discussions: Beautiful Symmetry and Ugly Symmetry [31]

To stress the important points of the present theory and to apply it in a relaxed manner to much more general situations, let us follow the style of discussing symmetry concept in a recent book [1b] and clarify the relation of symmetry with other concepts. Whether symmetry is beauty is of particular interest as the relationship of symmetry and beauty has been the subject of serious scientific research [32].

### *12.1. Order and Disorder*

According to common intuition, the standard of beauty is that beauty correlates with less chaos or more order which in turn correlates with more information and less entropy. Our discussions of the following will be based on this general correlation of beauty–orderliness–information. Therefore, a consideration of this problem may be reduced to a proper definition of order (table 1) and disorder.

**Table 1.** Two definitions of order.

| Definition | Order = periodicity or symmetry | Order = nonsymmetry and difference |
|---|---|---|
| Formation | Generated spontaneously | Not spontaneously |
| Examples | Chemical oscillation (symmetry and periodicity in time) or crystals (symmetry and periodicity in space) | Gas A stays in the left chamber and B in the right chamber to follow the order as shown in figure 5 |
| Reference | “Order Out of Chaos” [4] | The present work |
| Comment | Challenges the validity of the second law | Conforms to the second law |

However, unfortunately there are two totally different definitions of order or orderliness. Because they are opposite in meaning, it is only possible for us to conduct a meaningful discussion in chemistry and physics if only the suitable one of the two definitions is taken and consistently used. The first one regards "ordered" pattern as of the "order out of chaos" through "self-organization" [4]. Because the system is allowed to evolve spontaneously [4] the organized structure must be closer to an equilibrium state. It is well known that a system closer to equilibrium has a higher value of entropy of the universe or of an isolated system. It follows that a more "ordered" system has more entropy [4]. This kind of misconception is due to the correlation of higher entropy with lower symmetry because symmetry has been regarded as order. In this definition of order [4], order is the periodicity or symmetry generated spontaneously or through a "self"-action. This conclusion obviously violates the second law of thermodynamics [4], as commented by the present author [33].

In the other and the proper definition of order, order is generated and maintained by the surroundings (including the experimenter or information register and their equipment, etc.) [7d]. This order is achieved by applying confinements or constraints to the concerned system such as containers to restrict molecules spatially, or applying forces to produce distinguishability (and information). One example is shown in figure 5 where the two parts are confined and they are distinguishable. The social order is generated and maintained by applying discipline to its members to create distinguishability. Without discipline, if everyone in a country claims to be as powerful as the president and behaves in the same manner, the permutation symmetry is high and the society is in chaos. If two or more book locations are changed and the order of the library is unchanged, there is permutation symmetry in this library. This library either has many (to say 10000) copies of the same book, or is in a totally chaotic situation. Some years ago, when the present author did experiments using a refrigerated room where thousands of compounds prepared by the professor's former students were stored, he decided that symmetry was by no means order: If one brings any symmetry to the storeroom and accordingly reshapes the professor's nonsymmetrical arrangement of the precious samples (most of them are themselves asymmetric molecules) to "make order", the professor would react angrily. Now the present author is supervising a sample archive center [34] and has an even stronger picture that the archive order of the samples in a stock room is generated and maintained by confining samples to distinguishable containers and locations.

The statement that "a static structure (not necessarily a crystal), which is a frozen dynamic structure, is more orderly than a system of dynamic motion" is correct. However, when two *static* structures are directly compared, the saying that "a symmetric static structure has more order or is more orderly than a nonsymmetrical static structure" will be completely incorrect and misleading if disorder has been defined as entropy or information loss.

As mentioned in section 8.2 and 9, for stability reasons, a system applied for storing information always has symmetries for certain kinds of property (e.g., the periodicity of the sugar-phosphate backbone for DNA in example 6). In most cases information is recorded by using only one of several properties. The symmetric structure regarding other properties can be taken as formatting (section 2.3). The properly defined order or the genetic information in a DNA is recorded by the nonsymmetric structural feature (the 4 bases), not at all by the sugar or phosphate which are responsible for the symmetry. Strictly speaking, symmetry is not order; it is chaos.

*12.2. Symmetry and Diversity*

Intuitively, it can be easily understood that entropy and symmetry increase together if we simply use some common sense: it is indistinguishable (which is symmetry, see figure 6); and there is no information (or there is a large amount of entropy)! Whereas a similarity ($Z$) of a system can be defined as

$$Z = \frac{S}{L} = \frac{\ln w_S}{\ln w} \tag{19}$$

the diversity of the system ($D$) can be defined as

$$D = \frac{I}{L} = \frac{\ln w_I}{\ln w} \tag{38}$$

For more details, see ref. 7c-7e. Why are diversities [7c], such as molecular diversity, biodiversity, cultural diversity, etc. appreciated by many people, and why is diversity beautiful? The new theory gives a clear answer. If a collection of 10000 stamps has extremely high similarity, which means that all of them are of the same figure and size, even though this kind of stamp might be by itself very interesting, the permutation symmetry is obvious: If the positions of any two stamps are exchanged, the album of stamps remains the same. This collection is not beautiful and not precious because it lacks diversity. A library of 10000 copies of the same book is not nice either. Based on this idea, as a chemist, the present author initiated MDPI [34], an international project for collecting diverse chemical samples to build up molecular diversity. He has not collected 10000 samples of the same compound, even though it might be very interesting, e.g., the famous $C_{60}$ molecule. The molecule $C_{60}$ is itself beautiful not because of its symmetry, but because of its distinct structure and property compared to many other molecules and its contribution to the diversity of molecules we have studied.

For similar reason, it is true that we may feel a symmetric object beautiful if we have many other objects of less symmetry in our collection or in our experience. This symmetric object contributes to the diversity.

Diversity is beautiful. Symmetry is not. Coffee with sugar and cream is an interesting drink because of its diversity in taste (sweet and bitter) and color (white and dark). Diversity in a mixture makes the so-called high throughput screening of bioactivity testing possible [7c]. A country (e.g., USA or Switzerland) is considered a nice place because of its tolerance to all kinds of diversity (racial diversity, cultural diversity, religious diversity, etc.). Without such appreciation of diversity, this country would become much less colorful and less beautiful. If everyone behaves the same as you, looks the same as you, and there is a lot of symmetry, the world would be truly ugly. Democracy might be regarded as a sort of social or political diversity.

Unfortunately a system with high diversity is less stable. Synthetic organic chemists know that samples with impurities are less stable than highly purified samples. A mixture of thousands of compounds from a combinatorial synthesis is stable because the properties of the compounds are very similar and they normally belong to one type of molecule with the same parent structure. A mixture of acid and base is not stable. Storing a bottle of HCl and a bottle of $NH_3$ together is not a good idea.

A useful or beautiful system (a collection) should have a compromise between stability (due to indistinguishability or permutation symmetry and entropy) and diversity (due to distinguishability, nonsymmetry, diversity and information). As we have discussed before on complementarity (section 9 and the example), there can be many components or individuals that are complementary. A system satisfying this kind of complementarity is the most beautiful one with the most stability due to indistinguishability (symmetry) and the information due to distinguishability (nonsymmetry). A stable diversity is beautiful and practically useful for all kinds of diversity preservation [7c].

### *12.3. Ugly Symmetry*

#### 12.3.1. Assembling

Symmetry has been defined as beauty in many books and research papers [3, 8, 31]. According to the *Concise Oxford Dictionary*, symmetry is beauty resulting from the right proportion between the parts of the body or any whole, balance, congruity, harmony, keeping (page 1 of ref. 8).

Almost every modern arts museum collects a few paintings which are very symmetric, sometimes completely indistinguishable between the parts with the same pure white color (or gray or other color) overall (figure 6a). Symmetry is beautiful? OK, that's it! Why not? It is ugly because the emperor is topless and bottomless, even though many great intellectuals can cheerfully enjoy his most beautiful clothes (figure 6a). The most symmetric and most fussy paintings of those post-impressionists are the emperor's new clothes.

Crystals might be more beautiful than fluids because a solid structure has obviously lost all the dynamic symmetry such as isotropicity (see figure 6a). In a crystal, isotropicity does not exist: the properties along various directions become different (see figure 6b). This might be the reason why we like chocolate or ice cream in a certain form and may not like melted chocolate or runny ice cream soup.

#### 12.3.2. Individual Items

High speed motion around a fixed point and a fixed axis will create a figure of spherical and cylindrical symmetry, respectively. The electronic motion in a hydrogen atom around the proton creates a spherical shape for the hydrogen atom. These are many examples showing that the dynamic motion of a system results in an increased symmetry. These systems are stable but not necessarily beautiful. For example, a photo showing the face and shape of a girl figure skater with no symmetry is more beautiful than a photo of a mixture of the images facing many directions, or a mixture of structures at different angular displacements.

The famous molecule buckminsterfullerene $C_{60}$ is symmetric [1b] and very stable. We can predict that any modified $C_{60}$ mother structure with reduced symmetry will be less stable than the symmetric $C_{60}$.

Symmetric static structure is stable but not necessarily beautiful. A beautiful model, whether she stands or sits, never poses in a symmetric style before the crowd. A fit and beautiful body differs from the more symmetric, spherical shape of an over weight body. A guard in front of a palace stands in a symmetric way to show more stability (and strength). A Chinese empress sits in a symmetry style because stability is more interesting to her. The pyramid in Egypt is stable due to its symmetry. However, it is by no means the most beautiful construction.

Information will be lost if the ink is extremely faint or the same color as the background. A crystal is more stable than a noncrystal solid because the former has high static symmetry, a perfect symmetry without any information (figure 6b). Children understand it: on the walls of children's classrooms and bedrooms a visitor may find a lot of paintings, drawings and even scrawls done by the kids. If you do not put some colorful drawings there, children will create them themselves (e.g., figure 7b). The innocent children like to get rid of any ugly symmetry surrounding them. If your kids destroy symmetry on the walls, they are doing well.

Perfect symmetry is boring [32,1b], isn’t it? The very symmetric female face of oriental beauty [32] can be much more attractive if it is made less symmetric by a nonsymmetric hairstyle or by a nonsymmetric smile. The combination of certain symmetry (or symmetries) contributing stability according to our theory and certain nonsymmetry (or distinguishability, contributing certain interesting information) is the ideal beauty. The earth (figure 2) is beautiful because of its combination of symmetry and nonsymmetry. If it were of perfect symmetry and we could not distinguish North America, Europe or Asia, the world would be a deadly boring planet.

### *12.4. "Symmetry Is Beauty" Has Been Misleading in Science*

Beauty and its relation to symmetry has been a topic of serious scientific investigations [32]. This is fine. However, it has been widely believed that if the beholder is a physicist or a chemist, beauty also means symmetry. This may have already practically misled scientific research funding, publication and recognition. The situation in chemistry may be mentioned. Experimental chemists doing organic or inorganic synthesis may have more difficulty in synthesizing a specific structure of less symmetry than that of a highly symmetric one. However, a large number of highly symmetric structures can be much more easily published in prestigious journals (*Angewandt Chemie*, for example), because these molecular structures have been taken as more beautiful. Consequently other synthetic chemists have been regarded as intellectually lower achievers than those preparing highly symmetric structures. It has been shown by the history of chemical science and demonstrated by the modern arts of chemistry, particularly organic synthesis, that chemists endeavor to seek for asymmetry related to both space and time (nonsymmetry is the cause of phenomena [2]) not at all for symmetry [31]. The symmetric buckminsterfullerene $C_{60}$ is beautiful. However, many derivatives of $C_{60}$ have been synthesized by organic chemists. These derivatives are less symmetric, more difficult to produce and might be more significant. None of the drugs (pharmaceuticals) discovered so far are very symmetric. Very few bioactive compounds are symmetric. Because all of the most important molecules of life, such as amino acids, sugars, and nucleic acids, are asymmetric, we can also attribute beauty to those objects that are practically more difficult to create and yet practically more significant. The highest symmetry means equilibrium in science and death in life [31].

Sometimes, even graphic representation and illustration can be biased by the authors to create false symmetry. One example is shown in figure 14.

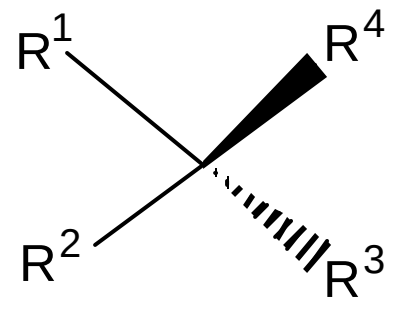

**4**

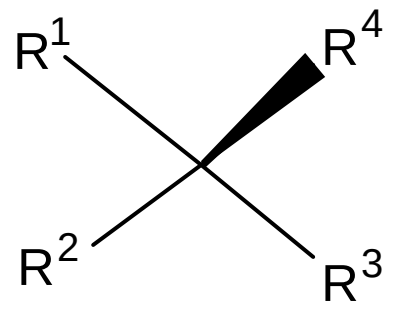

**5**

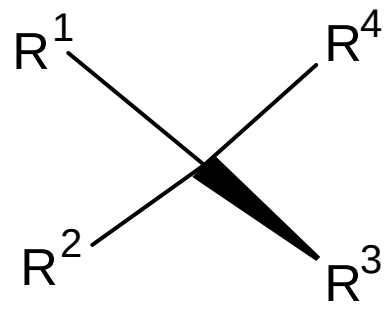

**6**

**Figure 14.** Stereochemical representation with false mirror symmetry in structure **4**: Both wedges have the thick ends at $R^3$ and $R^4$ placed identically away from the center [35]. The one-wedge representation (**5** or **6**) offers unambiguity and esthetic appeal [35a].

## 13. Conclusion

The structural stability criteria of symmetry maximization can be applied to predict all kinds of symmetry evolution. We have clarified the relation of symmetry to several other concepts, namely higher symmetry, higher similarity, higher entropy, less information and less diversity and they are all related to higher stability. This lays the very necessary foundation for further studies in understanding the nature of the chemical process.

*Acknowledgements:* The author is very grateful to Dr. Kurt E. Geckeler, Dr. Peter Ramberg, Dr. S. Anand Kumar, Dr. Alexey V. Eliseev and Dr. Jerome Karle and numerous other colleagues for their kind invitations to give lectures on the topics of ugly symmetry and beautiful diversity. Dr. Istvan Hargittai and Dr. J. Rosen kindly reviewed this paper.